\documentclass[prl,reprint,superscriptaddress,preprintnumbers,x11names]{revtex4-1}
\pdfoutput=1

\usepackage[utf8]{inputenc}
\usepackage{lmodern}
\usepackage[T1]{fontenc} 
\usepackage{microtype} 

\usepackage{xcolor}

\usepackage{hyperref}
\hypersetup{
	colorlinks=true,
	linkcolor=Blue4,
	citecolor=Red4,
	urlcolor=Green4,
	linktoc=page
}

\usepackage{enumerate}
\usepackage{graphicx}
\usepackage[percent]{overpic}
\usepackage{tikz}
\usepackage{subcaption}
\usepackage[many]{tcolorbox}
\usepackage{booktabs}
\usepackage{multirow}
\tcbset{highlight math style={enhanced,
  colframe=DeepSkyBlue4,colback=Honeydew1}}

\usepackage{amsmath,amssymb,slashed,mathbbol,mathtools}
\usepackage{amsthm}
\usepackage{slashed}

\DeclareSymbolFontAlphabet{\amsmathbb}{AMSb}%

\newcommand{\dd}{\mathrm{d}}

\newcommand{\e}{\mathrm{e}}

\newcommand{\be}{\begin{equation}}
\newcommand{\ee}{\end{equation}}
\newcommand{\bea}{\begin{eqnarray}}
\newcommand{\eea}{\end{eqnarray}}

\newcommand{\f}[2]{\frac{#1}{#2}}
\newcommand{\R}{\mathbf{R}}

\renewcommand{\Im}{\text{Im}}

\newcommand{\U}{\text{U}}
\newcommand{\SU}{\text{SU}}
\newcommand{\SO}{\text{SO}}

\newcommand{\USp}{\text{USp}}

\begin{document}

\title{Supersymmetric wormholes in String theory}

\date{\today}

\author{\href{mailto:dastesiano@hi.is}{Davide Astesiano} and \href{mailto:ffg@hi.is}{Fri{\dh}rik Freyr Gautason}}

\affiliation{Science Institute, University of Iceland, Dunhaga 3, 107 Reykjav{\'i}k, Iceland}

\begin{abstract}
\noindent We construct a large family of Euclidean supersymmetric wormhole solutions of type IIB supergravity which are asymptotically AdS$_5\times S^5$. The solutions are constructed using 
consistent truncation to maximally gauged supergravity in five dimensions which is further truncated to a four scalar model. Within this model we perform a full analytic classification of supersymmetric domain wall solutions with flat Euclidean domain wall slices. On each side of the wormhole, the solution asymptotes to AdS$_5$ 
dual to ${\cal N}=4$ supersymmetric Yang-Mills deformed by a supersymmetric mass term.

\end{abstract}


{\hypersetup{urlcolor=black}\maketitle}

\section{Introduction}
\label{sec:introduction}
The Euclidean path integral for quantum gravity is an important topic of research and for low-dimensional theories such as JT gravity, has recently lead to many fruitful results, see for instance \cite{Marolf:2020xie}. It has become clear that in low-dimensional theories it makes sense to sum over saddle points with different topologies \cite{Coleman:1988cy}. Still, in higher dimensions for standard Einstein-Hilbert gravity (coupled to matter) the rules remains somewhat unclear. Indeed, the story in higher dimensional gravity theories can be different from low-dimensional ones without leading to obvious inconsistencies.

In this regard, the role of wormholes, as possible saddle points of the path integral is still an important open problem \cite{Hebecker:2018ofv}. The processes that involve wormholes pose puzzles for unitarity of the quantum system and non-factorization of correlation functions in the holographic dual \cite{Kundu:2021nwp}. Moreover, the existence of  wormholes indicates that  probability amplitudes to produce or absorb baby universes are non-trivial which may lead to issues for  the Swampland program \cite{McNamara:2020uza,VanRiet:2020pcn}.

To improve our understanding it is necessary to provide the embedding of higher dimensional Euclidean wormholes in string theory. In this way, various ideas regarding the semi-classical formulation of gravity can be put to a test. Even more useful would be an explicit AdS/CFT realization in string theory that exhibits wormhole saddles. Research in this direction was initiated in some earlier works \cite{Witten:1999xp,Maldacena:2004rf}.

In order to construct Euclidean wormhole geometries we generally need a source of negative Euclidean energy. In string theory there is a natural way to obtain the required negative energy, which is to consider axion fields \cite{Giddings:1987cg}. When a Lorentzian theory containing axions is analytically continued to Euclidean, the axion kinetic term may become negative definite which gives rise to the required negative energy-momentum tensor. This was emphasized in \cite{Gutperle:2002km, Bergshoeff:2004pg, Bergshoeff:2005zf}, where it was pointed out that there are other scalar fields (saxions) with positive energy-momentum which pairs with the axions. In this way it is possible to have different kind of geometries, not only wormholes, depending on the sign of the overall energy-momentum tensor.

In the present work \footnote{A similar approach was taken in \cite{Buchel:2004rr} where wormholes in the model of \cite{Pilch:2000ue} were studied. The wormholes found there were neither supersymmetric nor stable.} we will consider a consistent truncation of type IIB supergravity on $S^5$ down to five-dimensional maximal supergravity coupled to $\SO(6)$ gauge group \cite{Gunaydin:1984qu,Pernici:1985ju,Gunaydin:1985cu}. The consistent truncation means that every solution of the five-dimensional model can be `uplifted' to a solution of full type IIB supergravity \cite{Pilch:2000ue,Lee:2014mla,Baguet:2015sma}. Our model will be a further (consistent) truncation to a four scalar theory coupled to AdS gravity originally introduced in \cite{Pilch:2000fu,Bobev:2016nua} to study the holographic duals to ${\cal N}=4$ Supersymmetric Yang-Mills (SYM) deformed by a mass parameter. Recall that ${\cal N}=4$ SYM can be deformed by three independent mass parameters while retaining ${\cal N}=1$ supersymmetry. The mass deformed theory is usually denoted by ${\cal N}=1^*$ and we will use that language here. The four scalar supergravity model we use assumes that the three mass parameters in the dual field theory are all equal leading to an $\SO(3)$ flavor symmetry. We will see that the same model gives rise to singular domain walls, as well as regular Euclidean wormholes that have much in common with the original axionic wormholes of \cite{Giddings:1987cg}. 

The embedding of axionic wormholes as AdS compactifications of 10D (or 11D) supergravity has been discussed recently in the literature  \cite{ArkaniHamed:2007js,Hertog:2017owm, Ruggeri:2017grz,Astesiano:2022qba,Loges:2023ypl}. In short, the existence of the wormhole solution relies on the fact that the supergravity theory can be truncated such that the moduli space of scalars give rise to the Lagrangian
\be\label{toy}
\mathcal{L} = R -\frac{1}{2} \mathcal{G}_{IJ}\partial \phi^I\partial\phi^J - 2\Lambda\,.
\ee
Here we denote the AdS moduli (massless scalar fields) by $\phi^I$, $\mathcal{G}_{IJ}$ is the non-trivial metric on the moduli space, and $\Lambda$ is the negative cosmological constant. As mentioned before, in Euclidean supergravity the moduli space metric is not necessarily positive definite. When we Wick rotate to Euclidean signature in space-time, the axions get flipped signs \footnote{Which is consistent with their target space metric shift symmetry.}, while the other scalar fields remains untouched \cite{Burgess:1989da, Bergshoeff:2005zf}.


In the model studied in this paper, we will encounter a similar feature when rotating to Euclidean signature. Only one of the four scalars is a modulus which happens to be the five-dimensional dilaton. The dilaton is related to the Yang-Mills coupling constant in the dual ${\cal N}=4$ SYM theory. Furthermore, the constant cosmological constant $\Lambda$ in \eqref{toy} is replaced by a non-trivial potential that depends on all but one scalar field
\begin{align}
    2\Lambda \rightarrow {\cal P}(\phi^I)\,.
\end{align}
It is therefore clear that $\phi^I$ are not moduli anymore. This added feature will make the equations of motion much more complicated than for standard axionic wormholes. However with the help of first order BPS equations that ensure supersymmetry of the solutions, we will be able to find supersymmetric Euclidean wormholes. In more detail, the solution we construct is a five-dimensional domain wall of the form
\be\label{metricansatzintro}
\dd s_5^2 = \dd r^2 + \e^{2A(r)} \dd s_{\R^4}^2\,,
\ee
where $\dd s_{\R^4}^2$ is the flat metric on $\R^4$ and the metric function $A$ only depends on $r$. The $\R^4$ can be compactified to $T^4$ when considering the Euclidean gravity path integral. The solution we find will be described by a metric function $A(r)$ that approaches the standard asymptotic AdS form $A(r) \sim \pm r/2L$ for both $r\to \pm\infty$ connected by a region of smooth region of non-AdS space. Close to the asymptotic boundary we find a pair of (Euclidean) AdS$_5$ spaces dual to ${\cal N}=4$ SYM. The scalar fields and the metric have asymptotic form which is consistent with masses for the three chiral fields in ${\cal N}=4$ being turned on. Supersymmetry is therefore broken from ${\cal N}=4$ to ${\cal N}=1$ at the asymptotic boundary and indeed the full geometry preserves ${\cal N}=1$ supersymmetry. It is interesting to note that in general the asymptotic behaviour of the two boundaries is not exactly the same beyond leading order. This can be interpreted as the two QFTs dual to each side of the wormhole are both ${\cal N}=1^*$, but with different Yang-Mills coupling constants and different VEVs turned on. A special line of solutions exists where the configuration is slightly more symmetric and the two boundary theories have the same VEVs. On this line there is a very special point where the neck of the wormhole shrinks to zero size and the metric becomes singular. The singular solution encountered there is the very well known GPPZ solution \cite{Girardello:1999bd}\footnote{Strictly speaking, the GPPZ solution found here is the one for which the (dimensionless) gaugino vev takes a maximal value. In the notation of \cite{Pilch:2000fu,Bobev:2019wnf} this is the $\lambda=1$ solution.}.

An important question when faced with  wormhole solutions such as these ones is whether they dominate over the corresponding `disconnected geometry'. Since we have been focusing on supersymmetric solutions, the disconnected geometry should also preserve supersymmetry. In our analysis we have been able to fully classify solutions to the BPS equations subject to the metric ansatz \eqref{metricansatzintro}. It turns out that for a given set of boundary conditions which allows for a wormhole solution, there is no corresponding disconnected solution. Disconnected solutions could perhaps be found by relaxing some of the isometries built into the metric ansatz \eqref{metricansatzintro}, but we have not carried out a general analysis.
It is straight-forward to check that for our BPS solutions,  the regularized on-shell Euclidean action vanishes 
\begin{align}
    S_{\text{E}}=0\,.
\end{align}

As discussed in the literature (e.g. \cite{Maldacena:2004rf}), wormhole solutions are usually subject to various kinds of instabilities such as tachyonic directions in field space. 
Since our solutions are supersymmetric, we expect them to be protected from the standard instabilities and therefore we do expect that our solutions contribute to  the euclidean path integral. Without the disconnected geometries it is impossible for us to determine at this stage whether or not the wormholes dominate.

\section{5D supergravity}
 \label{sec:5d}
The supergravity model considered here is a four scalar truncation of maximal 5D supergravity with $\SO(6)$ gauge group \cite{Gunaydin:1984qu,Gunaydin:1985cu,Pernici:1985ju}. The 5D $\SO(6)$ gauged maximal supergravity has been shown to arise as consistent truncation of type IIB supergravity on $S^5$ \cite{Pilch:2000ue,Lee:2014mla,Baguet:2015sma} and so any solution of the maximal supergravity can be embedded into type IIB supergravity. 

The four scalar truncation discussed presently was first introduced in the holographic study of the ${\cal N}=1$ mass deformation of ${\cal N}=4$ SYM with all three mass parameters taken to be equal \cite{Pilch:2000fu,Bobev:2016nua,Bobev:2019wnf}. When all masses are equal the QFT possesses $\SO(3)$ flavor symmetry which (if we assume it is not spontaneously broken) can be utilized on the supergravity side to truncate the maximal theory such that the bosonic sector contains a metric and eight scalar fields \cite{Pilch:2000fu}. A further discrete symmetry can be imposed to truncate the theory even further leaving only four scalar fields apart from the metric.

As the name suggests, the scalars of the model parametrize a four-dimensional subspace of the full 42 dimensional scalar manifold $E_{6(6)}/\USp(8)$. This subspace consists of two copies of the Poincar{\'e} disc
\be
{\cal M} = \Big(\f{\SU(1,1)}{\U(1)}\Big)^2\,,
\ee
which we parametrize with two complex scalar fields $z_1$ and $z_2$.
The five-dimensional supergravity action of the four scalar model takes the form
\be
S = \f{1}{16\pi G_N} \int \star\Big(R-2 K_{i\bar\jmath}\dd z^i\cdot\dd {\bar z}^{\bar\jmath}-{\cal P}\Big)\,,
\label{action1}
\ee
where the scalar potential is
\be
{\cal P} =  {\f12}\e^{K}\Big(K^{i\bar\jmath}D_i{\cal W} D_{\bar\jmath}\overline {\cal W} - \f83 |{\cal W}|^2\Big)\,,
\ee
with the K{\"a}hler covariant derivative defined as $D_i f= (\partial_i +\partial_i K)f$, and the K{\"a}hler metric defined by $K_{i\bar\jmath} = \partial_i\partial_{\bar\jmath}K$ and its inverse is $K^{i\bar\jmath}$. We have written the theory in terms of the K{\"a}hler potential $K$ and a holomorphic superpotential ${\cal W}$ which are given by \footnote{Note that we have changed slightly conventions as compared to e.g. \cite{Bobev:2019wnf}. Explicitly
\be
z_1^\text{\cite{Bobev:2019wnf}}=\f{z_1-i}{z_1+i}\,,\quad z_2^\text{\cite{Bobev:2019wnf}}=\f{-z_2+i}{z_2+i}\,,\quad {\cal P}= 4{\cal P}^\text{\cite{Bobev:2019wnf}}\,.
\ee}
\be
K = -\sum_{i=1}^2 \log(2\Im z_i)\,,\qquad 
{\cal W} =  3gz_2(z_1+z_2)\,.
\ee
The theory exhibits the scaling symmetry
\be\label{scaling}
z_i \mapsto \lambda z_i\,,
\ee
which leaves the action invariant. This is nothing but the dilatonic shift symmetry, but the five-dimensional dilaton is indeed one of the four scalar fields present in the model. As we will see later on, this symmetry implies a conserved quantity which is helpful when the equations of motions are solved.

The maximally supersymmetric vacuum solution of the maximal five-dimensional supergravity is obtained as a critical point of this model by setting
\be\label{vacuum}
z_1=z_2=i\e^{\varphi}\,,
\ee
where $\varphi$ is the constant value given to the five-dimensional dilaton. For the vacuum solution, the scalar potential takes the value ${\cal P} = -3g^2$ and therefore the metric is AdS$_5$ with length scale $L = \f{2}{g}$.


\section{BPS equations and Wormhole solutions}


We are interested in finding flat sliced supersymmetric domain wall solutions to the equations of motion of the four scalar model. To this end we assume that the five-dimensional metric takes the form \eqref{metricansatzintro} and assume that all scalar fields as well as the metric function $A$ are only functions of the radial variable $r$. A supersymmetric solution must satisfy the first order equations \cite{Bobev:2019wnf}
\be\label{BPSformal}
{\cal E}_A \equiv A' + \f13 W =0\,,\quad {\cal E}^i \equiv (z^i)' - K^{i\bar\jmath}\partial_{\bar\jmath}W=0\,,
\ee
where the real superpotential $W$ is defined as $W = \e^{K/2}|{\cal W}|$. We have verified that all solutions to the BPS equations are also solutions to the five-dimensional equations of motion. It should be noted at this point that in Lorentzian supergravity $\bar z_{\bar\imath}$ is the complex conjugate of $z_i$ and the same holds true for ${\cal W}$ and $\overline{\cal W}$. In this paper we will also consider Euclidean solutions where $\bar z_{\bar\imath}$ is best treated as independent from $z_i$. In general $z_i$ and $\bar z_{\bar\imath}$ still represent 2 real degrees of freedom in total. This feature has been discussed previously in e.g. \cite{Bobev:2016nua} but will become more apparent later when we discuss the explicit solution to the BPS equations.

In order to simplify the system of BPS equations, we introduce new field variables
\be
z_1 = i\e^{\varphi+3\alpha + i\theta_1}\,,\qquad z_2 = i\e^{\varphi-\alpha + i\theta_2}\,.
\ee
The new scalar fields have a clear interpretation from the perspective of the holographic dual field theory. In particular, $\varphi$ is the five-dimensional dilaton and is dual to the marginal Yang-Mills coupling. $\alpha$ is dual to a scalar bilinear operator transforming in the ${\bf 20}'$ representation of $\SO(6)$ and $\theta_{1,2}$ are dual to two fermion bilinear operators transforming in the ${\bf 10}\oplus\overline{\bf 10}$ representation. As we will see momentarily a special solution of the system of BPS equations \eqref{BPSformal} can be obtained by taking $\varphi$ and $\alpha$ to be trivial. These solutions were initially found and analyzed in \cite{Girardello:1999bd} and later uplifted to ten dimensions in \cite{Petrini:2018pjk,Bobev:2018eer}.

It turns out to be useful to further define new sets of variables 
\be
t_1 = \tan\theta_1\,,\quad t_2 = \tan\theta_2\,,
\ee
in order to eliminate most of the trigonometric functions. Even with these new variables the BPS equations are quite lengthy and difficult to analyze. In order to make progress we replace the field $\alpha$ by a new variable $X$ defined by
\be
X\equiv \f1{2(1+t_2^2)}\Big( 1+ t_1t_2+\sqrt{1+t_1^2}\sqrt{1+t_2^2}\cosh 4\alpha \Big)\,.
\ee
This definition of $X$ may seem ad hoc at first but it is closely related to the real superpotential $W$. Nevertheless the explicit form we use for $X$ was identified by trial and error in order to obtain as simple BPS equations as possible. With these definitions the BPS equations take the form
\be
\begin{split} \label{BPSrvariable}
4\sqrt{X}(t_1') & = 3g \Big(t_2-t_1+2X t_1(1+t_2^2)\Big)\,,\\
4\sqrt{X}(t_2') & =   g  \Big(t_1-t_2+6X t_2(1+t_2^2)\Big)\,,\\
4\sqrt{X}(X') &= 8gX(X-1)\,,\\
4\sqrt{X}(A') &= -2gX  (1+t_2^2)\,.
\end{split}
\ee
Writing the BPS equations in these coordinates has simplified them significantly enabling us to fully solve them. Note that the dilaton has been decoupled completely from the system as it does not appear on the right-hand side of any of the equations. Indeed the dilaton does not appear in the scalar potential and can thus be deterimined by simple integration once the other scalar fields are found. This does not mean that the dilaton is constant however as its BPS equation has a complicated right-hand side. We will come back to the dilaton after we have solved the system \eqref{BPSrvariable}.

Recall now that the dilaton scaling symmetry \eqref{scaling} present in our model implies the existence of a constant of motion \footnote{Identifying this constant of motion is quite tricky because naively the Noether charge associated with \eqref{scaling} seems to vanish identically. However by slightly deforming the metric ansatz and considering curved domain walls the correct conserved quantity \eqref{jdef} can be derived \cite{futurewormholes}.}
\be\label{jdef}
j = -\f{g^3}{64}\e^{3A}(t_1+3t_2)\,,
\ee
which implies that we do not have to solve explicitly the equations for both $t_1$ and $t_2$, only one combination of them suffices. For this purpose we identify another combination of the $t$-scalars 
\be\label{Ydef}
Y \equiv \f{g^3}{64}\e^{3A}(t_1-t_2)\,,
\ee
whose BPS equation is particularly simple
\be
\sqrt{X}(Y') = -gY\,.
\ee
At this point it is straightforward to integrate the BPS equation for $X$. It turns out to be slightly more convenient to write down the solution using $Y$ as the coordinate, but then we have
\be\label{Xsol}
\f{\dd X}{\dd Y} = \f{2X(1-X)}{Y}\,,\quad \text{or}\quad X = \f{Y^2}{k+Y^2}\,,
\ee
where $k$ is a real integration constant. Due to the fact that we are considering a flat domain wall with a priori non-compact slices, we are free to shift the metric function $A$ by a constant and absorb the shift into coordinate redefinition along the four-dimensional slices. Since the definition of $Y$ \eqref{Ydef} involves $\e^{3A}$ this shift of $A$ implies a rescaling of $Y$ which in the solution for $X$ in \eqref{Xsol} implies that $k$ can be rescaled by an arbitrary real constant. This means that without loss of generality we can consider three distinct values $k=\{-1,0,1\}$.
Next we use that
\be
j+Y =  -\f{1}{16}g^3\e^{3A}t_2\,,
\ee
to write
\be \label{Adiff}
\f{\dd A}{\dd Y} = \f{1}{2} \f{Y}{k+Y^2} \Big(1+256 g^{-6}\e^{-6A}(j+Y)^2\Big)\,.
\ee
Finally, we remark that in the $Y$-coordinate, the five dimensional metric takes the form
\be
\dd s_5^2 = \f{\dd Y^2}{g^2(Y^2+k)} + \e^{2A(Y)}\dd s_{\R^4}^2\,.
\ee
Note that the solution is invariant under the simultaneous change of sign
\be\label{signflips}
j\mapsto -j\,, \quad Y\mapsto -Y\,,\quad t_{1,2}\mapsto -t_{1,2}\,.
\ee

We will now solve the equation \eqref{Adiff} for the three cases $k=\{-1,0,1\}$. For $k=0$ we find
\be
\f{g^6}{2^6}\e^{6A} = 4j a Y^3-12 Y^2-12jY-4j^2\,,
\ee
where $a$ is a real integration constant. For non-zero $ja$, the resulting five dimensional metric approaches the AdS$_5$ vacuum solution for $jaY\gg 1$. $\e^{6A}$ is a cubic polynomial of $Y$ with a discriminant $\Delta \sim -j^2(1+j^2a)^2$ that is negative definite or zero. If the discriminant is non-zero, $\e^{6A}$ has exactly one root where the metric is singular. If $j=0$, then $\e^{6A}$ has one single root and one double root, the double root is a maximum. Coming from asymptotic infinity the metric function is a decreasing function and therefore the metric always crosses the single root first and is singular. Finally if $1+j^2a=0$, then $\e^{6A}$ has a triple root where once again the metric is singular.

When $k=\pm1$ the metric can be expressed as
\be
\f{g^6}{2^6}\e^{6A} = 4\Big(2jk Y^3 -3Y^2-j^2-2k+2a(Y^2+k)^{3/2}\Big)\,.
\label{ASol}
\ee
where we have recycled $a$ as the integration constant. The AdS$_5$ asymptotics are reached for $|Y|$ large if $jk\,\text{sgn}(Y)+a>0$. Hence it is possible here that there are two AdS$_5$ asymptotic regions if $a>|j|$. However for $k=-1$, there is always a singularity present in the center (either when $|Y|$ reaches 1 from above or earlier). For this reason and since we are interested in wormhole solutions, we focus on $k=1$ from now on.

A wormhole solution is obtained if the metric function $\e^{6A}$ has two AdS$_5$ asymptotic regions (for $|Y|$ large) and is otherwise positive. Since $\e^{6A}$ has at most two real roots, we have to ensure that the discriminant of the polynomial $(2jk Y^3 -3Y^2-j^2-2k)^2-4a^2(Y^2+k)^{3}$ is negative implying it has no real roots. Combined with the condition that $a>|j|$ we find wormhole solutions if and only if
\be\label{wormholecond}
1+\f{j^2}{2}< a\,.
\ee
It is now easy to see that the scalar field $\alpha$ is purely imaginary when the above condition is satisfied. In fact the condition for $\alpha$ being real is that
\be\label{realitycond}
1+j^2\ge a^2\,.
\ee
The boundary of \eqref{realitycond} is where the scalar $\alpha$ vanishes throughout and can be identified with the GPPZ solutions \cite{Girardello:1999bd}\footnote{In order to find the precise map to the GPPZ solution we must identify the integration constant $a$ with $\pm\f{1+\lambda}{2\sqrt{\lambda}}$ where $\lambda$ is the integration constant as used in e.g. \cite{Bobev:2019wnf}.}. The two regions, defined by \eqref{wormholecond} on one hand and \eqref{realitycond} on the other, are completely non-overlapping. It is interesting to note that the GPPZ solution with $j=0$ and $a=1$ (or $\lambda=1$ in the notation of \cite{Bobev:2019wnf}) is infinitesimally close to being a wormhole and can be viewed as the limiting solution where the wormhole neck shrinks to zero size. This particular solution was previously emphasized in \cite{Bobev:2019wnf} as playing a special role in the family of GPPZ solutions as it is the unique vacuum that is selected when the four scalar system is placed on the sphere \cite{Bobev:2016nua}.

\begin{figure}[ht]
\centering
\includegraphics[width=0.4\textwidth]{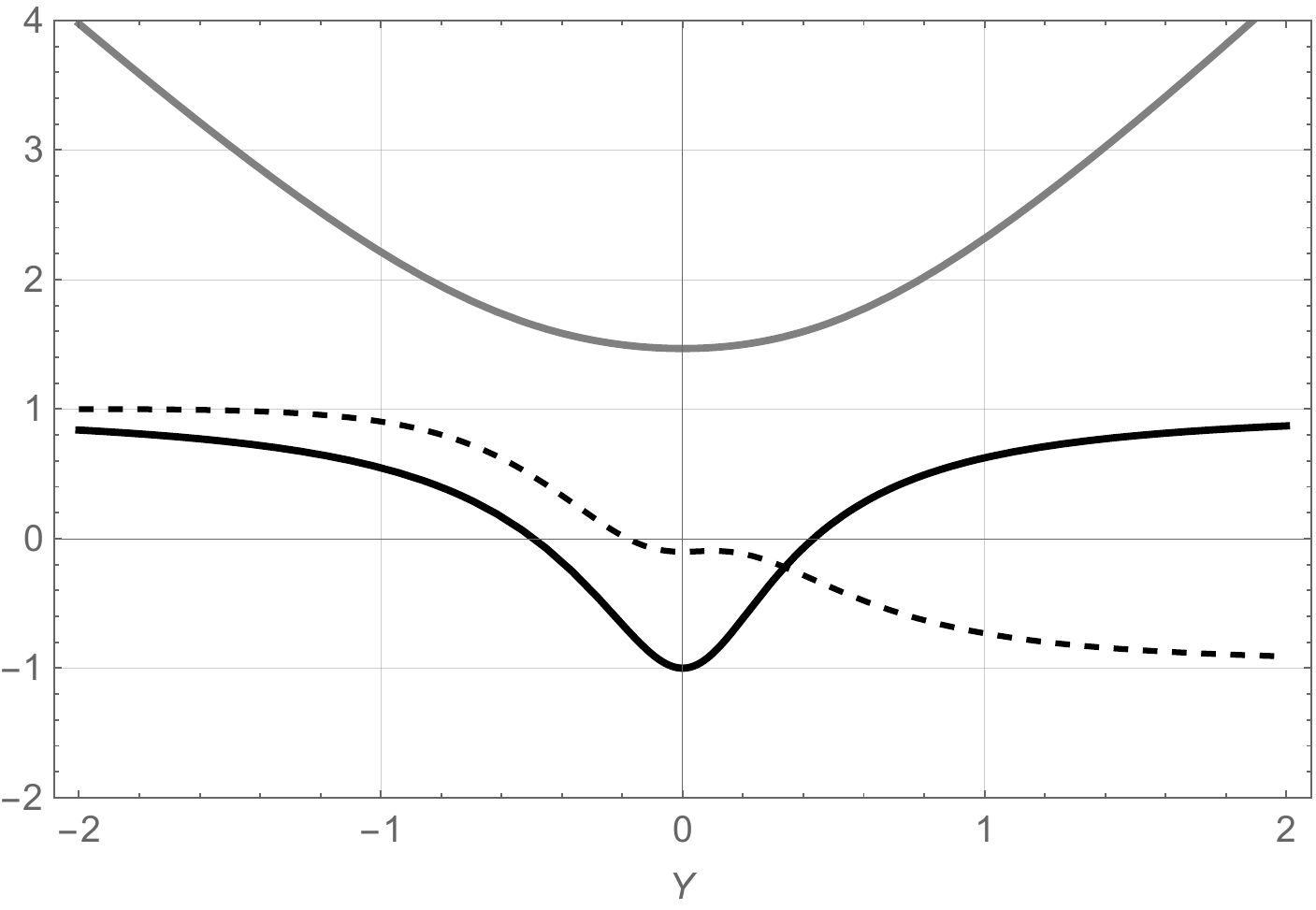}
\caption{\label{wormholeplot}A plot of a sample wormhole solution for $a=1.4$ and $j=0.1$. The metric function $g^2\e^{2A}/4$ is drawn in grey, $\cosh4\alpha$ is drawn in solid black and $\cosh4\varphi$ in dashed black.}
\end{figure}

In addition to the $\alpha$ scalar being imaginary, the dilaton is also imaginary. In order to see that we have to integrate the BPS equation for the dilaton which for $k=1$ takes the form
\begin{widetext}
\be\label{dilatonsol}
\varphi(Y)-\varphi_0=
- \int_{-\infty}^Y\frac{3 y (y-j) \sqrt{\left(y^2+1\right)^3
   \left(-a^2+j^2+1\right)}}{2 \left(y^2+1\right)^2 \left(a
   \sqrt{y^2+1}+j y-1\right) \left[y^2 \left(a \sqrt{y^2+1}+3\right)+a
   \sqrt{y^2+1}+j \left(y^2-3\right) y-1\right]}\,\dd y\,.
\ee
\end{widetext}
We have not been able to perform the integral analytically but it is easy to do numerically. We display a plot of a sample solution in Fig. \ref{wormholeplot}. A generic feature of the wormhole solutions is that they are not symmetric around $Y=0$, even for $j=0$. This is clearly observed from Fig. \ref{wormholeplot} where the dilaton is far from being symmetric around $Y=0$. Furthermore, the symmetry \eqref{signflips} indicates that the scalars $t_1$ and $t_2$ are antisymmetric around $Y=0$ for $j=0$.

The fact that both $\alpha$ and $\varphi$ are imaginary may appear problematic. But as we have already mentioned in the introduction and will discuss in more detail in the conclusion, this is the expected behaviour for Euclidean supergravity. Indeed when $S^4$ domain wall solutions were constructed in \cite{Bobev:2013cja,Bobev:2016nua} the same feature was expected and observed.

\section{field theory interpretation and on-shell action}
Recall that the scalar fields $\alpha$, $\varphi$, and $\theta_{1,2}$ have a direct relation to operators in the dual field theory. More precisely if $\theta_1 = -3\phi_1 + \phi_2$ and $\theta_2 = \phi_1+\phi_2$, and we think of ${\cal N}=4$ SYM in ${\cal N}=1$ language, then $\phi_2$ is dual to the gaugino bilinear and $\phi_1$ is dual to the three chiral multiplet fermion bilinear. The latter are all equal since we assumed that the $\SO(3)$ flavor symmetry preserved by the equal mass ${\cal N}=1^*$ Lagrangian is not spontaneously broken. By performing a expansion of the fields around the AdS$_5$ asymptotics we can identify the sources and vevs given to the dual operators for our solution. A general UV expansion compatible with the BPS equations takes the form
\be\label{generalUVexp}
\begin{split}
A &= -\f12 \log \epsilon - \f{m^2}{2} \epsilon +{\cal O}(\epsilon^2)\,,\\
\phi_1 &= m\epsilon^{1/2} -\f{5}{6}m^3\epsilon^{3/2}+ {\cal O}(\epsilon^2)\,,\\
\phi_2 &= w\epsilon^{3/2} + {\cal O}(\epsilon^2)\,,\\
\alpha &= v\epsilon + {\cal O}(\epsilon^2)\,,\\
\varphi &= \varphi_0  + {\cal O}(\epsilon^2)\,.
\end{split}
\ee
Here $\epsilon$ is the small parameter controlling the distance from the asymptotic boundary. From this expansion we see that the parameter $m$ is the mass given to the chiral fields wheras $w$ is the gaugino vev and $v$ is the so-called chiral condensate, i.e. the vev of the scalar bilinear in the chiral multiplets.

For the wormhole solutions with $k=1$, we have two asymptotic regions that are located at $Y\to \pm\infty$. It is a straight-forward task to expand our solution around these asymptotic regions and match onto the general expansion in \eqref{generalUVexp} to identify the relation between the physical parametrs of the ${\cal N}=1^{*}$ theory and the integration constants $j$ and $a$. In this way we find the dimensionless quantities
\be
\f{w_\pm}{2m^3} =(j^2+1)\pm a\, j\,,\qquad \f{v^2}{m^4} = (j^2+1) - a^2 \,,
\label{Asymptotics}
\ee
where we have denoted the two gaugino condensates that are encountered in the two asymptotic regions $Y\to\pm \infty$ by $w_\pm$. The dimensionless chiral condensate, $v/m^2$, is the same in both regions. As a consequence of the condition \eqref{wormholecond}, regular wormhole solution exist only when the chiral condensate $v/m^2$ is imaginary. Finally it should be noted that the asymptotic value for the dilaton always differs between the two asymptotic regions. We do not have an analytic expression for the difference between the two asymptotic values but this is easy to work out on a case-by-case basis using the expression \eqref{dilatonsol}. The conclusion is that the Euclidean wormhole solution is a bulk geometry that connects two copies of mass-deformed ${\cal N}=4$ SYM where some of the boundary conditions (including the Yang-Mills coupling constant) are different.

An important aspect for the evaluation of the gravitational partition function, and of the free energy of the dual theory, is the on-shell action of our wormholes solutions.
Adding the Gibbons-Hawking term to the action and performing a partial integration, the action can be rewritten in terms of the squared BPS equations
\begin{align}
    \mathcal{L}+\mathcal{L}_\text{GH}= - 
   e^{4A} \left[12{\cal{E}_A}^2 - 2 K_{i\bar\jmath} {\cal{E}}^i \overline{\cal{E}}^{\bar\jmath}\right] - 2 \partial_r (e^{4A} W), \nonumber
\end{align}
where $W$ is the real superpotential we defined before $W = \e^{K/2}|{\cal W}| $.
Evaluating this on-shell, the BPS equations set to zero the first two terms, leaving only the total derivative. The total derivative term leads to a divergent expression which must be regulated. As explained in \cite{Freedman:2013oja,Bobev:2016nua} the correct supersymmetric counter term (when the holographic boundary is flat) should be chosen to exactly cancel the total derivative, i.e. 
\begin{align}
S_\text{c.t.}=  \f{1}{8\pi G_N}\int_{\partial M} e^{4A} W\,.
\end{align}
This implies that for all supersymmetric regular wormhole solutions found in this paper, the on-shell action vanishes.

\section{Final Comments}
As anticipated in the introduction, our wormhole solutions are supported by a negative term in the energy-momentum tensor. In fact, the dilaton $\varphi$ and the field $\alpha$ are imaginary while preserving the reality of the metric as well as the action. This is interpreted as a Wick rotation on target space which must be simultaneously performed when Wick rotating space-time.
In terms of the physical coordinates, the target space metric takes the form
\begin{align}
   -2 K_{i\bar\jmath}\dd z^i\cdot\dd {\bar z}^{\bar\jmath}&= \sec ^2\theta _1 \big(\left(3 \dd\alpha+\dd\varphi \right){}^2+\dd\theta _1{}^2\big)+ \nonumber\\
   &+ 3 \sec
   ^2\theta _2\big(\left(\dd\alpha-\dd\varphi \right){}^2+\dd\theta _2{}^2\big) \label{Targetspace1}\,.
\end{align}
The target space metric exhibits a pair of translation symmetries
\begin{align}
\alpha\rightarrow\alpha+ \text{const}\,,\quad \varphi\rightarrow\varphi+ \text{const}\,.
\end{align}
From the perspective of the target space metric the scalars $\alpha$ and $\varphi$ appears as axions. However it must be noted that the scalar potential depends non-trivially on $\alpha$ and so the shift `symmetry' of $\alpha$ is not a true symmetry of the theory. Nevertheless, according to the prescription in \cite{Hertog:2017owm,Astesiano:2022qba}, the Wick rotation of Lorentzian supergravity to Euclidean should be accompanied with a similar Wick rotation in target space
\begin{align}
    \alpha\rightarrow i \alpha\,,\quad \varphi\rightarrow i \varphi\,.
\end{align}
While the space-time signature becomes Euclidean, the target space signature is now Lorentzian. This is not particularly surprising if we remember that in Euclidean signature, the R-symmetry of ${\cal N}=4$ SYM is not $\SO(6)$ but rather $\SO(1,5)$. 
At the same time, the Wick rotation also affects the potential $V$. As discussed before, the dilaton does not appear in the potential and the Wick rotation of $\alpha$ does not spoil the reality of the potential. This can be directly observed from the expression for the scalar potential in the $t_{1,2}$, $X$-variables:
\be
{\cal P} = -\frac{3}{4} g^2 \left(2-2(t_2^4-1)X-t_2(t_1-3 t_2)t_2\right)\,.
\ee
We see here that even though $\alpha$ is imaginary, both $t_{1,2}$ and $X$ are real and so the potential is indeed real.

The supergravity theory we studied  in this work is a particular truncation of the  five-dimensional ${\cal N}=8$ supergravity. Within this truncation we completely classified the flat supersymmetric domain wall solutions leading in particular to the wormhole solutions discussed here. It is reasonable to expect that this class of wormholes is but a small subset of all possible wormhole solutions in ${\cal N}=8$ supergravity. As an example, the four scalar truncation studied here embeds in a much larger ten-scalar truncation \cite{Bobev:2016nua} where we expect to find an even larger class of Euclidean wormhole solutions once the scalar target space has been Wick rotated appropriately. This and further study of domain wall solutions in the ten-scalar model is a subject of future work \cite{futurewormholes}. This approach, although not directly ten-dimensional, yields Euclidean wormholes in type IIB supergravity rather cleanly, as any solution of maximal $\SO(6)$ gauged five-dimensional supergravity can be uplifted to a solution of type IIB supergravity on $S^5$ \cite{Pilch:2000ue,Lee:2014mla,Baguet:2015sma}. It would be interesting to uplift our wormhole solutions to ten-dimension and analyze them further there.

Due to the similarity between four- and five-dimensional ${\cal N}=8$ supergravities, one may wonder whether similar wormhole solutions exist also in four dimensions. $\SO(8)$ gauged four-dimensional supergravity is a consistent truncation of 11-dimensional supergravity on $S^7$ and therefore a Euclidean wormhole solution would uplift directly to a wormhole in 11D. 

Our findings seem to contradict the analysis conducted in appendix D of \cite{Marolf:2021kjc} where  preliminary investigation did not yield regular wormholes within this truncation. 
It is not clear to us whether there are some assumptions made in \cite{Marolf:2021kjc} which our solutions do not satisfy. We allowed the scalar fields to assume imaginary values 
which was a crucial ingredient that enabled us to successfully identify and construct regular Euclidean wormholes.

Since our solutions preserve supersymmetry and are regular we do not expect any instabilities to arise and question the validity of them. By all accounts they should then contribute to the Euclidean path integral. Since we did not find disconnected geometries with the same boundary conditions as the wormholes, we are unable to answer whether the wormholes dominate or not. If they dominate, then it leads to the well-known factorization puzzle in holography \cite{Witten:1999xp,Maldacena:2004rf} in this case  for deformations of AdS$_5$ dual to the ${\cal N}=1^*$ theory. How this puzzle is resolved is a open question at this stage. One possibility is that fermion zero-modes in the spectrum cause the wormhole contribution to the path integral to vanish. This was indeed observed in \cite{Iliesiu:2021are} and in that case it was related to supersymmetry being broken. Our backgrounds only partially break supersymmetry and so it remains an open question whether fermion zero-modes prevent the Euclidean wormholes found here from making a non-trivial contribution.

%
\section*{Acknowledgements}
We are grateful to Nikolay Bobev, Valentina G. M. Puletti, Krzysztof Pilch, Thomas Van Riet,  Watse Sybesma, and L{\'a}rus Thorlacius for useful discussions. We thank Nikolay Bobev, Thomas Van Riet, and L{\'a}rus Thorlacius for comments on the manuscript.
FFG and DA are supported by the Icelandic Research Fund under grant 228952-052. FFG is partially supported by grants from the University of Iceland Research Fund.

\bibliography{references}

\end{document}